\documentclass[twocolumn,amsmath,amssymb,superscriptaddress]{revtex4}

\pdfoutput=1

\usepackage[pdftex]{graphicx}
\usepackage[pdftex,colorlinks=true,urlcolor=blue,linkcolor=black]{hyperref} 

\usepackage{graphicx}
\usepackage{dcolumn}
\usepackage{bm}
\usepackage{datetime}

\usepackage{enumitem}
\setitemize{leftmargin=*}

\begin{document}

\title{From Few to Many: Observing the Formation of a Fermi Sea One Atom at a Time} 

\author{A.\,N. Wenz$^{1,2 \, *\, \dagger}$, G. Z\"urn$^{1,2 \,\dagger}$, S. Murmann$^{1,2}$, I. Brouzos$^{4}$, T. Lompe$^{1,2,3}$, and S. Jochim$^{1,2,3}$\\
\normalsize{$^{1}$Physikalisches Institut, Ruprecht-Karls-Universit\"at Heidelberg, 69210 Heidelberg, Germany.}\\
\normalsize{$^{2}$Max-Planck-Institut f\"ur Kernphysik, Saupfercheckweg 1, 69117 Heidelberg, Germany}\\
\normalsize{$^{3}$ExtreMe Matter Institute EMMI, GSI Helmholtzzentrum f\"ur Schwerionenforschung, 64291 Darmstadt, Germany.}\\
\normalsize{$^{4}$Institut f\"ur Quanten-Informationsverarbeitung, Universit\"at Ulm, 89069 Ulm, Germany.}\\
\normalsize{$^\ast$To whom correspondence should be addressed. E-mail:  wenz@physi.uni-heidelberg.de}\\
\normalsize{$\dagger$These authors contributed equally to this work.}
}

\date{\today }

\begin{abstract}
\textbf{Knowing when a physical system has reached sufficient size for its macroscopic properties to be well described by many-body theory is difficult.
We investigate the crossover from few to many-body physics by studying quasi one-dimensional systems of ultracold atoms consisting of a single impurity interacting with an increasing number of identical fermions. We measure the interaction energy of such a system as a function of the number of majority atoms for different strengths of the interparticle interaction. As we increase the number of majority atoms one by one we observe the fast convergence of the normalized interaction energy towards a many-body limit calculated  for a single impurity immersed in a Fermi sea of majority particles. }
\end{abstract}

\maketitle
The ability to connect the macroscopic properties of a many-body system to the microscopic physics of its individual constituent particles is one of the great achievements of physics. This connection is usually made using the assumption that the number of particles tends to infinity. Then a transition from discrete to continuous variables can be made, which greatly simplifies the theoretical description of large systems. When does a system become large enough for this approximation to be valid? This is a difficult question to answer because most calculations based on a microscopic description become prohibitively complex before their predictions approach the many-body solution. 
Experimentally, this question has been studied for example in the context of Helium droplets \cite{Grebenev1998} and nuclear physics \cite{Migdal1959} by measuring the emergence of superfluidity for increasing system size. We address this question using ultracold atoms which have already been used to study few-particle systems with tunable interactions \cite{Ospelkaus2006b,Will2010,Will2011}.

In our experiment we control the system size on the single particle level while maintaining full control over the interparticle interactions. We achieve this by deterministically preparing few-particle systems of ultracold  $^6$Li atoms \cite{Serwane2011}, whose interparticle interaction can be tuned using Feshbach resonances \cite{Chin2010, Bergeman2003}. This allows us to explore the crossover from few to many-body physics by studying the fermionic quantum impurity problem, where a single impurity atom interacts with a number of fermionic majority atoms. This has the advantage that the majority atoms do not  interact with each other due to the Pauli principle. In this system the impurity acts as a test particle which we use to probe the majority component. The limit of a single majority particle (Fig. 1A) has been thoroughly investigated both theoretically \cite{Busch1998} and experimentally \cite{Ospelkaus2006b, zuern2012}. For a large number of majority atoms the system can be described as a single impurity interacting with a Fermi sea (Fig. 1C). Similar physics has been experimentally explored by introducing a small fraction of impurity atoms into a large Fermi sea of ultracold atoms \cite{schirotzek2009, Koschorreck2012, kohstall2012}.

\begin{figure} [tbh]
\centering
	\includegraphics [width= 5.5cm] {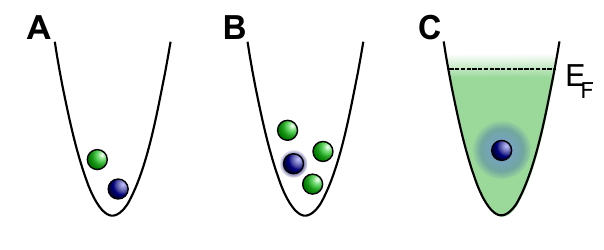}
	\caption{\textbf{From few to many.} A single impurity (blue) interacting with one, few and many fermions (green) in a harmonic trapping potential. In the many-body case the majority component can be described as a Fermi sea with a Fermi energy E$_{\mathrm{F}}$.}	
	\label{fig:figure1}
\end{figure}

\begin{figure*} [tbh]
\centering
	\includegraphics [width= 17.5 cm] {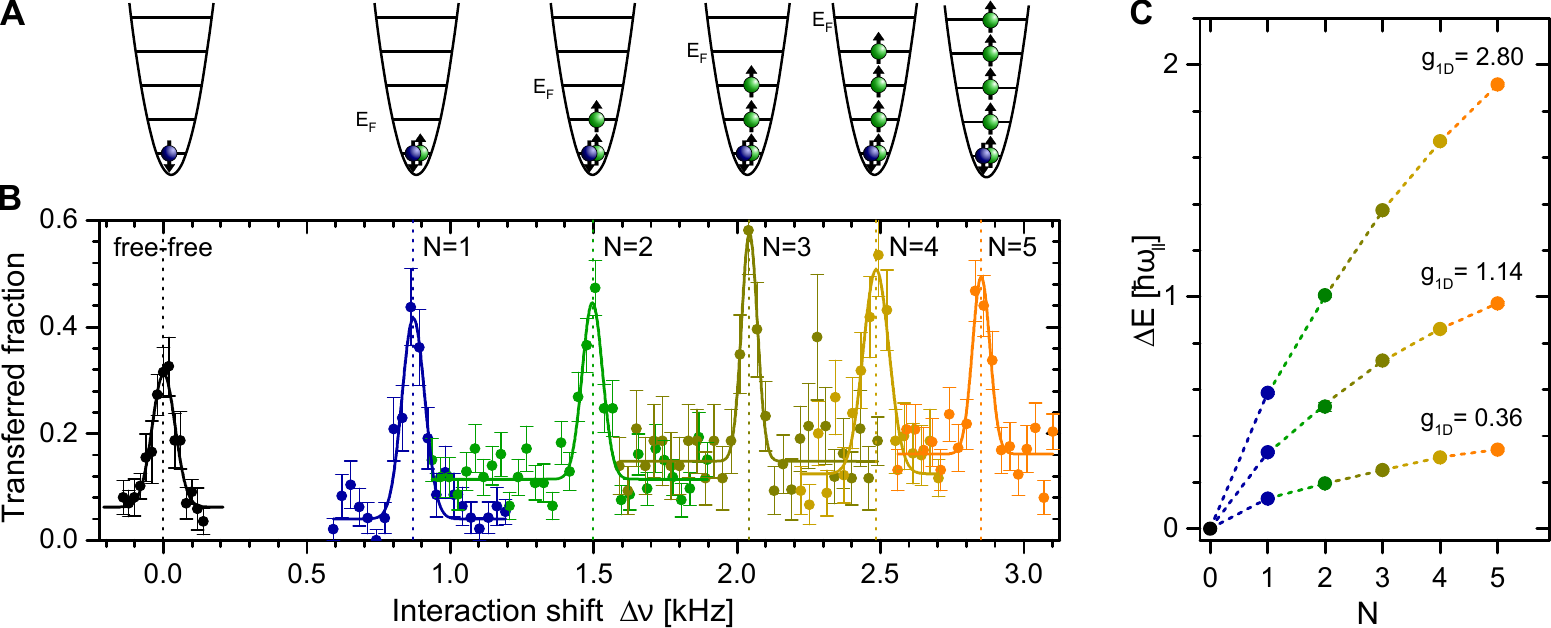}
	\caption{\textbf{Interaction energies.} \textbf{(A)} Deterministically prepared non-interacting $(N+1)$-particle states. \textbf{(B)} 
	Determined radio-frequency (RF) spectra for a single impurity interacting with different numbers of majority atoms with a coupling strength $g_{\text{1D}}= 2.80 $ in units of $a_{\parallel} \, \hbar \, \omega_{\parallel}$. The transferred fraction denotes the average fraction of the minority atom transferred to a different hyperfine state.\textbf{(C)} The interaction energies for different values of $g_{\text{1D}}$ 
	are determined by Gaussian fits to the experimental spectra (solid lines in B) and plotted as a function of $N$. The errors of $\Delta E(N)$ are determined from the uncertainties of the fits to the experimental spectra \cite{SOM} and are not visible due to their small size.}
	\label{fig:figure2}
\end{figure*}

We prepare $(N+1)$-particle systems consisting of one impurity atom and $N$ majority atoms in their $N$-particle ground state trapped in an elongated optical dipole potential (Fig. 2A) \cite{Serwane2011}. 
We create these quasi one-dimensional few-particle systems using ultracold fermionic $^6$Li atoms in different hyperfine states, where we label the minority particle as $\mid\downarrow \rangle$ and the atoms of the majority component as $\mid\uparrow \rangle$. This system is very well described by the Hamiltonian
\[
	H = \sum_{i=0}^{N}{\left( - \frac{\hbar^2}{2m}  \frac{\partial^2}{\partial x_i^2}  +  \frac{1}{2} m \omega_{\vert\vert}^2 \, x_i^2  \right)}   +  
	  g_{\text{1D}} \sum_{i=1}^{N} \delta(x_i -x_{0} ) .  
\] 
Here $x_0$ is the position of a single impurity atom that interacts with $N$ atoms of the majority component, which are located at the coordinates  $x_1$ to $x_N$, through a contact interaction with the strength $g_{\text{1D}}$, $\hbar$ is the reduced Planck's constant, $m$ is the mass of a $^6$Li atom and $\omega_{\vert\vert}$ is the axial frequency of the harmonic trapping potential \cite{SOM}.
This Hamiltonian can be solved analytically for the two limiting cases of just one majority particle ($N=1$) \cite{Busch1998} and for an infinite number of majority particles in a homogeneous system ($\omega_\parallel \rightarrow 0)$ \cite{Mcguire1965}.

To probe the system we measure the interaction energy between the impurity and the majority component as a function of the number of majority particles.  
We do this by changing the internal state of the single impurity atom from an initial hyperfine state $\mid\downarrow_i\rangle$ to a different hyperfine state $\mid\downarrow_f\rangle$ with a radio-frequency (RF) pulse \cite{julienne2005, zuern2013}. If no majority atoms are present, the transition occurs at a frequency $\nu_0$ corresponding to the hyperfine splitting between the initial and final states of the impurity atom. In the presence of $N$ majority atoms the interactions between the impurity and the majority atoms lead to an interaction shift $\Delta\nu(N)$ of the transition frequency. 

Using such RF spectroscopy measurements we study systems with weak ($g_{\text{1D}} = 0.36$), intermediate ($g_{\text{1D}} = 1.14$) and strong ($g_{\text{1D}}= 2.80$) repulsive interactions, where $g_{\text{1D}}$ is given in units of $a_{\parallel} \, \hbar \, \omega_{\parallel}$ with the harmonic oscillator length $a_{\parallel}=\sqrt{(\hbar/(m \omega_\parallel)}$.
We perform these measurements for systems containing between one and five majority atoms. As an example, the resulting RF spectra for strong interaction are shown in Fig. 2B. Here one should note that a trapped system has discrete eigenstates and thus all RF spectra consist of discrete transitions. As the resolution of our RF spectroscopy is much higher than the level spacing $\hbar \omega_{\parallel}$ of our trap we resolve these discrete transitions and therefore observe sharp transitions with symmetric line shapes. Hence we can fit all these transitions with Gaussians and obtain the interaction energies $\Delta E(N)=h \; \Delta\nu(N)$ for different coupling strengths $g_{\text{1D}}$ (Fig. 2C).

The addition of each majority atom increases the number of particles the impurity atom can interact with and thus the interaction energy rises for increasing $N$. 
For weak interactions it was shown that $\Delta E\propto \sqrt{N}$  \cite{Astrakharchik2013} and hence $\Delta E$ diverges for $N \to \infty$. Therefore, we rescale the interaction energy by the natural energy scale of the system, the Fermi energy of the majority atoms $E_F$, to obtain the dimensionless interaction energy $\mathcal{E}=\Delta E/E_F$. 
In a trapping potential the Fermi energy is determined by the energy of the lowest single particle level not occupied by a majority atom. As we only consider the gain in interaction energy, we neglect the zero-point motion, which means that $E_F= N \hbar \omega_\parallel$ for the case of a harmonic trap. 
For our optical trap, which is slightly anharmonic, we determine the Fermi energy for each $N$ by precise measurements of the level spacings and WKB calculations \cite{zuern2012,SOM}.

\begin{figure*} [tbh]
\centering
	\includegraphics [width= 12cm] {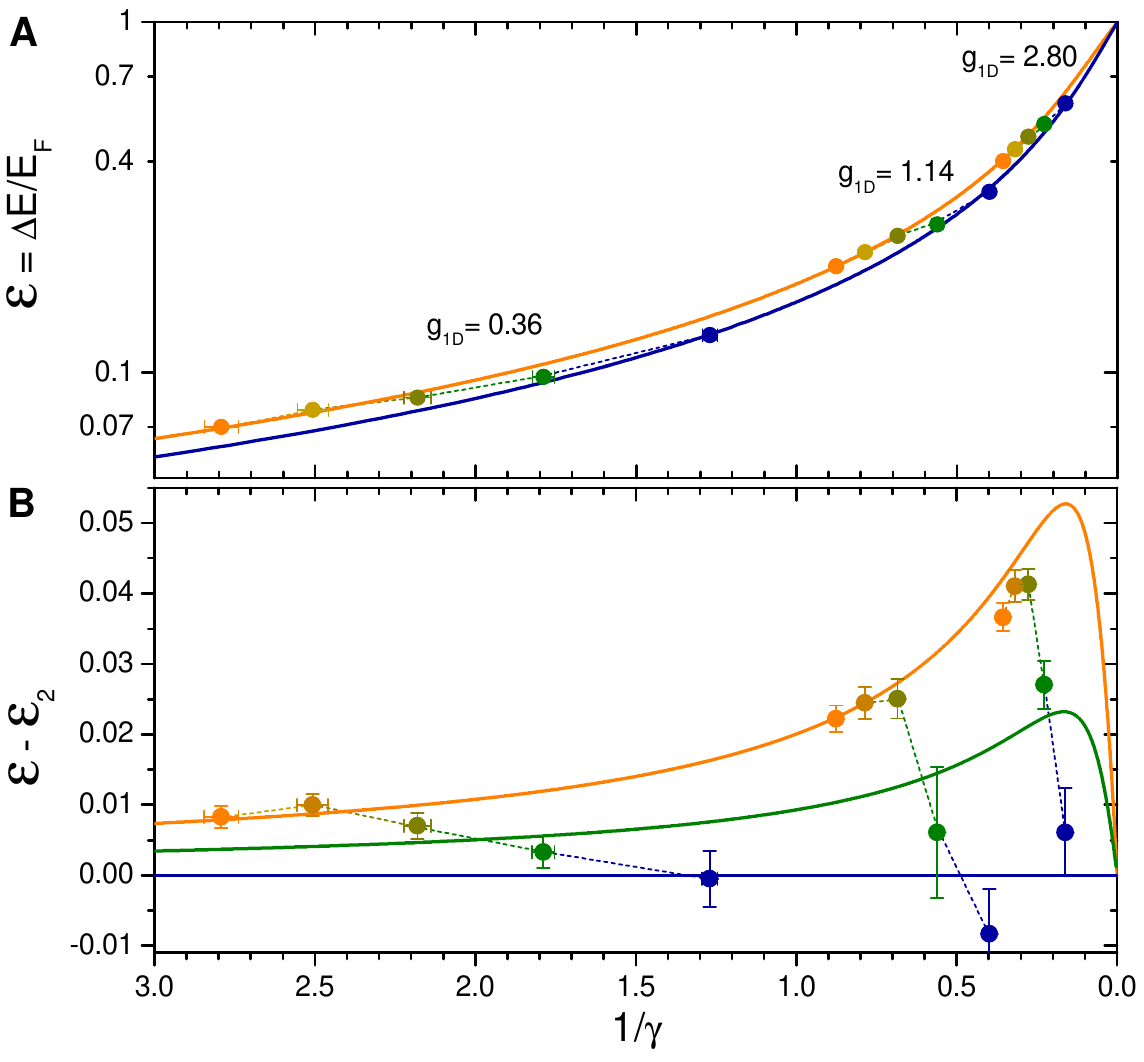}
	\caption{\textbf{Comparison with theory.} \textbf{(A)} Normalized interaction energy as a function of the interaction parameter $\gamma$. The solid lines indicate the analytic predictions for $N=1\,$\cite{Busch1998} (blue) and $N\rightarrow \infty\,$\cite{Mcguire1965} (orange). The colored dots represent the measured interaction energies for systems with $N= 1\dots5$  (blue to orange), corrected for the effects of the finite aspect ratio of the trap. For the normalization of the interaction energies of the few-particle theories and the measurement we have approximated the density by its peak density to compare it with the many-body theory developed for a homogeneous system \cite{Mcguire1965}. \textbf{(B)} Difference between the interaction energies and the theoretical prediction for a two-particle system ($N=1$). The theoretical prediction for $N=2$\,\cite{Gharashi2012} is shown as a green solid line. The measured interaction energies coincide with the two-particle prediction for $N=1$ and for $N\geq4$ approach the many-particle limit which indicates the formation of a Fermi sea.
 }
		\label{fig:figure3}
\end{figure*}

To compensate for the change in density caused by an increase in the number of particles we rescale $g_{\text{1D}}$ with the line density of the majority atoms. For the trapped systems we approximate the line density by its peak density given by the Fermi momentum $k_F = \sqrt{2 m E_F}/ \hbar$, which has been shown to be very accurate even for low particle numbers \cite{Astrakharchik2013}.
As a result we obtain a dimensionless interaction parameter $\gamma=\frac{\pi  m}{\hbar^2} \, g_{\text{1D}}/k_F $ \cite{footnotedimensionlessparameter}.

To investigate how the measured interaction energy of our system approaches the many-body limit we compare our data with theoretical predictions for the two limiting cases of one and an infinite number of majority particles.
The normalized interaction energy $\mathcal{E}_2=\Delta E_2/E_F$ of the two-particle system consisting of an impurity atom interacting with just one majority atom \cite{Busch1998} is shown as a blue line in Fig. 3.
In the many-body limit where a single impurity is immersed in a Fermi sea of an infinite number of majority atoms an explicit expression for the interaction energy in a homogeneous one-dimensional system has been analytically calculated \cite{Mcguire1965} to be

\begin{equation}
 \mathcal{E}_{\infty}=\frac{\Delta E_{\infty}}{E_F}
=\frac{\gamma}{\pi^2}
\left[1 -  
\frac{\gamma}{4}+\left(\frac{\gamma}{2\pi}+\frac{2\pi}{\gamma}\right) \tan^{-1} \left( \frac{\gamma}{2\pi}\right)
 \right]. \nonumber
\end{equation}
This interaction energy for the many-body case is shown in Fig. 3 as an orange line.

The two theories coincide for the limits of vanishing and diverging interactions. For $\gamma = 0$ the interaction energy is zero. For $\gamma \to \infty$ one reaches the limit of fermionization where the energy of the impurity atom interacting with $N$ majority fermions exactly matches the energy of $N+1$ non-interacing identical fermions  and therefore the gain in interaction energy is equal to the Fermi energy $E_F$ \cite{Girardeau1960, Mcguire1965, Weiss2004, Haller2009, Girardeau2010, zuern2012, brouzos2012,zinner2013}.
In between these limits the many-body solution consistently has a slightly larger normalized interaction energy than the two-body solution.

To compare our data to these strictly one-dimensional theories we have to consider the fact that our quasi one-dimensional trapping potential has a finite aspect ratio of about $10:1$. We estimate this effect on the interaction energy using numerical calculations \cite{Idziaszek2004, Gharashi2012,Zuern-thesis} and find that the largest corrections occur for intermediate and strong interactions, and is on the order of 2\% of the interaction energy \cite{SOM}. The corrected data is plotted in Fig. 3A as a function of the dimensionless interaction parameter $\gamma$ where the different colors indicate the number of majority atoms. The measured interaction energy quickly approaches the many-body limit as the number of majority particles increases. To show this in more detail we subtract the interaction energy given by the two particle theory $ \mathcal{E}_2$ from the total interaction energy (Fig. 3B).

For the two-particle system, where the impurity interacts with only one majority atom ($N=1$) we find excellent agreement between the experimental data (blue dots) and its analytical prediction (blue line). For two or more majority atoms ($N\geq2$) there is no analytic prediction for the interaction energy but there are recent state-of-the-art numerical calculations \cite{Gharashi2012,Astrakharchik2013,conduit2013}. We compare our experimental results for $N=2$ (green dots) to the calculations of \cite{Gharashi2012} (green line) and find good agreement. Note that adding only one atom to the two-particle system leads to an increase of the normalized interaction energy which is already about half of the energy gain expected for the $N \to \infty$ limit.
 
For more than two majority particles one observes a quick convergence of the experimentally determined interaction energy towards the many-body limit described by the analytical solution in \cite{Mcguire1965}.
For strong interactions the data points follow the shape of the many-body theory but have a slight offset. Possible reasons for this deviation might be the peak density approximation and the uncertainties in our corrections for the anharmonicity and the finite aspect ratio, which are largest for strong interactions. For weak and intermediate interactions the determined interaction energy agrees with the many-body theory within the experimental errors for as few as four majority atoms. This fast convergence of the normalized interaction energy shows that already for a few particles the many-body theory describes essential aspects of our system.

To gain further insight into the one-dimensional quantum impurity problem, it would be interesting to consider our findings in terms of polaron-like behavior. In general a polaron is a quasiparticle that consists of an impurity coherently dressed by particle-hole excitations of the Fermi sea \cite{Prokofev2008a,Prokofev2008b,Giraud2009}. In a one-dimensional system one does not expect the existence of polaronic quasiparticles but can still find polaron-like behavior for weak interactions \cite{Guan2013}. To understand how the this behavior emerges for growing $N$ requires further measurements of the effective mass which can be obtained by investigating the excitation spectrum of the system.

For strong interactions the polaron-like description breaks down in a one-dimensional system and for $g_{1D} \to \infty$ one approaches the limit of fermionization. Then the system's energy reaches that of a system of $(N+1)$ non-interacting identical fermions \cite{footnotenumericalconstant}. By crossing this point of fermionization one can reach a metastable state where the interaction energy is larger than the Fermi energy \cite{zuern2012} which might lead to the appearance of nontrivial spin correlations \cite{conduit2013,Gharashi2013,zinner2013}. For spin balanced systems with attractive interactions our setup could be used to study the emergence of pairing \cite{Zurn2013_attractive} and superfluidity in a fermionic few-particle system with tunable interactions and full control over the system's quantum state.



\section*{Acknowledgments}
We thank D. Blume and S.E. Gharashi for inspiring discussions and for providing the results of their numerical calculations. We also thank R. Schmidt, X. Guan and J.McGuire for very helpful discussions. The authors gratefully acknowledge support from the ERC starting grant $279697$, the Helmholtz Alliance HA$216/$EMMI and the Heidelberg Center for Quantum Dynamics. A.N.W.  and  G.Z. acknowledge support by the IMPRS-QD. I.B. acknowledges G.E. Astrakharchik for helpful discussions and the FC funded FET project QIBEC for financial support.

\cleardoublepage

\section*{SUPPLEMENTAL MATERIAL}

\subsection*{Confining potential and coupling strength}
The confining potential for trapping our few fermion systems is created by the focus of a single laser beam with a wavelength of $1064\,$nm. This leads to a cigar-shaped trapping potential whose 
parameters have been precisely determined in \cite{zuern2012} by measurements of the level spacings between the lowest trap levels and WKB calculations. In a harmonic approximation we obtain trap frequencies of $\omega_\perp= 2 \pi \times 14.22\,(0.35)\,$kHz \cite{sala2013} in radial direction and $\omega_\parallel= 2 \pi \times 1.488\,(0.014)\,$kHz \cite{zuern2012} in axial direction.
For a harmonic trap with such an aspect ratio of $\omega_\parallel$:$\omega_\perp  \approx\,$1:10 a two-particle system can be well described by mapping it onto a true 1D system \cite{Idziaszek2004}.
The value of the corresponding 1D coupling constant $ g_{\text{1D}}$ is determined by the 3D scattering length $a_{\text{3D}}$ and by the harmonic oscillator length of the radial confinement $a_\perp=\sqrt{\hbar/(m \omega_\perp)}$ as 
\begin{equation}
 g_{\text{1D}}=\frac{2\hbar^2a_{3D}}{m a_{\perp}^2}\frac{1}{1-Ca_{3D}/(\sqrt{2}a_{\perp})},  \nonumber
\end{equation}
where $C$$=$$-\zeta(\frac{1}{2})$=$1.46$... and $\zeta$ is the Riemann zeta function \cite{Bergeman2003} .
In our experiment we tune $g_{\text{1D}}$ by changing $a_{\text{3D}}$ using a magnetic offset field  \cite{zuern2013}. A plot of $g_{\text{1D}}$ as a function of the magnetic field is shown in Fig. S2, where
 $g_{\text{1D}}$ is given in units of $a_{\parallel} \, \hbar \, \omega_{\parallel} $, which are the characteristic length and energy scales of our trap
\cite{footnoteg1d}.

\subsection*{Preparation of the initial sample}
For our experiments we use ultracold $^6$Li atoms in the three lowest-energy Zeeman sublevels of the electronic ground state which we label $\vert 1 \rangle$, $\vert 2 \rangle$ and $\vert 3 \rangle$ (see Fig. S1).
For most of our measurements we use initial systems where the majority component consists of atoms in state $\vert 3 \rangle$ and an impurity atom in state $\vert 1 \rangle$. To create these systems we start by preparing non-interacting  imbalanced few fermion samples of a single atom in state $\vert 1 \rangle$ and $N=1\dots5 $ atoms in state $\vert 2\rangle$ as described in \cite{Serwane2011} with a preparation fidelity of $\gtrsim 88\%$. To convert these into $\vert 1 \rangle$-$\vert 3\rangle$ samples we then perform an RF Landau-Zener passage which drives the atoms from  state $\vert 2 \rangle$ to state $\vert 3 \rangle$. This has a transfer efficiency of $95\%$ per atom which results in a combined preparation fidelity of the non-interacting $\vert 1 \rangle$-$\vert 3 \rangle$ sample $\gtrsim 68\%$. 

To introduce interactions between the minority atom and the majority component we ramp the magnetic field away from the zero crossings of $g_{\text{1D}}$ to the values given in table S4. 
At these larger interactions we observe some loss of atoms due to relaxation to lower energy states which increases with coupling strength and atom number \cite{zuern2013, conduit2013}.
To obtain an upper bound for the loss probability we measure the loss in a sample with $N=6$ and obtain a loss probability $\lesssim 29\%$ for strongest interaction.
Therefore we obtain a lower bound for the combined fidelity to prepare an interacting ($N+1$)-particle system of $\gtrsim 48\%$.
For systems with $N>5$ the preparation fidelity of the initial sample is considerably lower and the stability of the sample is significantly reduced which inhibits reliable results of the RF spectroscopy.

\begin{figure}[htbp]
	\begin{minipage}[b]{8.5cm}
	\includegraphics[width=\textwidth]{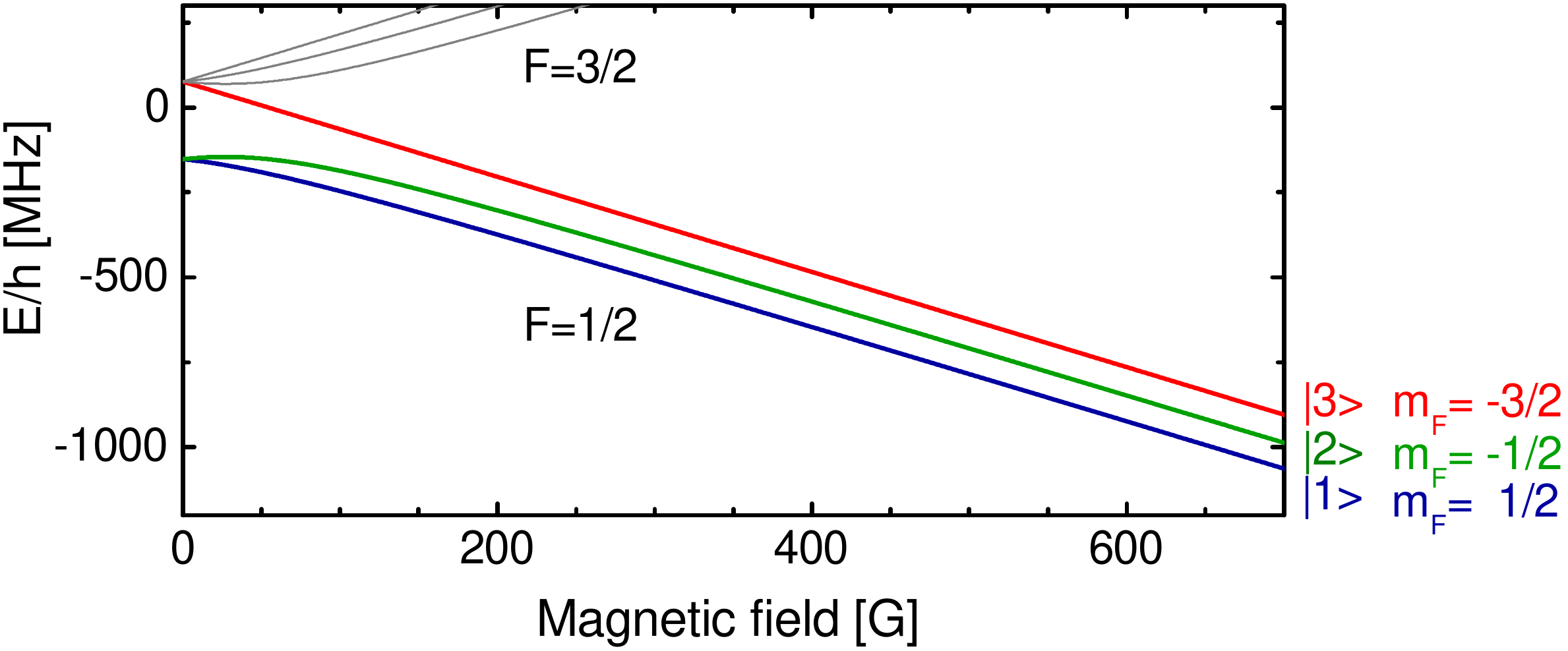}
	\label{fig:SOMfig1}
	\end{minipage}
	\begin{minipage}[b]{8.5cm} 
	\begin{itemize}
\item[]{Figure S1: Energy of the lowest Zeeman hyperfine sublevels of $^6$Li in the electronic ground state as a function of the magnetic offset field. The states are labeled according to their Zeeman energy starting from the lowest. For our experiments we use different combinations of atoms in states $\vert 1 \rangle$, $\vert2\rangle$ and $\vert 3 \rangle$.}
\end{itemize}
	\hfill
	\end{minipage}
\end{figure}

\subsection*{Spectrum and line shape in a trapped system}
In the presence of a confining potential the energy eigenstates of the
system are discrete. Therefore the spectrum consists of discrete transitions between the different energy levels. However if the resolution $\nu_\text{res}$ is lower than the difference between close-lying levels the excitation spectrum appears as a continuous spectrum (see Fig. S2 A). Examples for this regime -- where almost all RF measurements with ultracold atoms have been performed so far -- can be found in \cite{kohstall2012, Koschorreck2012,schirotzek2009}. In this case the measured RF spectrum has a characteristic line shape whose form depends on the excitation spectrum of the system.
Fig. S2 B illustrates the case where $\nu_{\text{res}}$ is comparable to the level spacing which has been achieved in \cite{zuern2013}.
In our experiment we are in the sideband resolved regime where the RF resolution is much higher than the level spacing and the transitions between different trap states are well separated (see Fig. S2 C and D).
Therefore the determined interaction energies are insensitive to saturation effects which only affect the relative height of the different trap sidebands.
By precisely measuring the relative height of these sidebands one could in future experiments determine for example the effective mass of the polaron-like state.

\subsection*{RF spectroscopy measurement}
To determine the interaction energy of our systems we use the fact that it is possible to change the hyperfine state of our atoms with an RF-field whose frequency is resonant with the energy splitting of the initial and final state. For a non-interacting system (i.e. a system with no majority atoms) this transition frequency is the Zeeman splitting of the two states given by the Breit-Rabi formula (see Fig. S1) which we label the free-free transition frequency $\nu_0$. 
If the initial or final system are interacting the transition is shifted by the difference of the interaction energies of the initial and final state. 

\begin{figure}[htb!]
	\begin{minipage}[b]{7.5cm}
	\includegraphics[width=\textwidth]{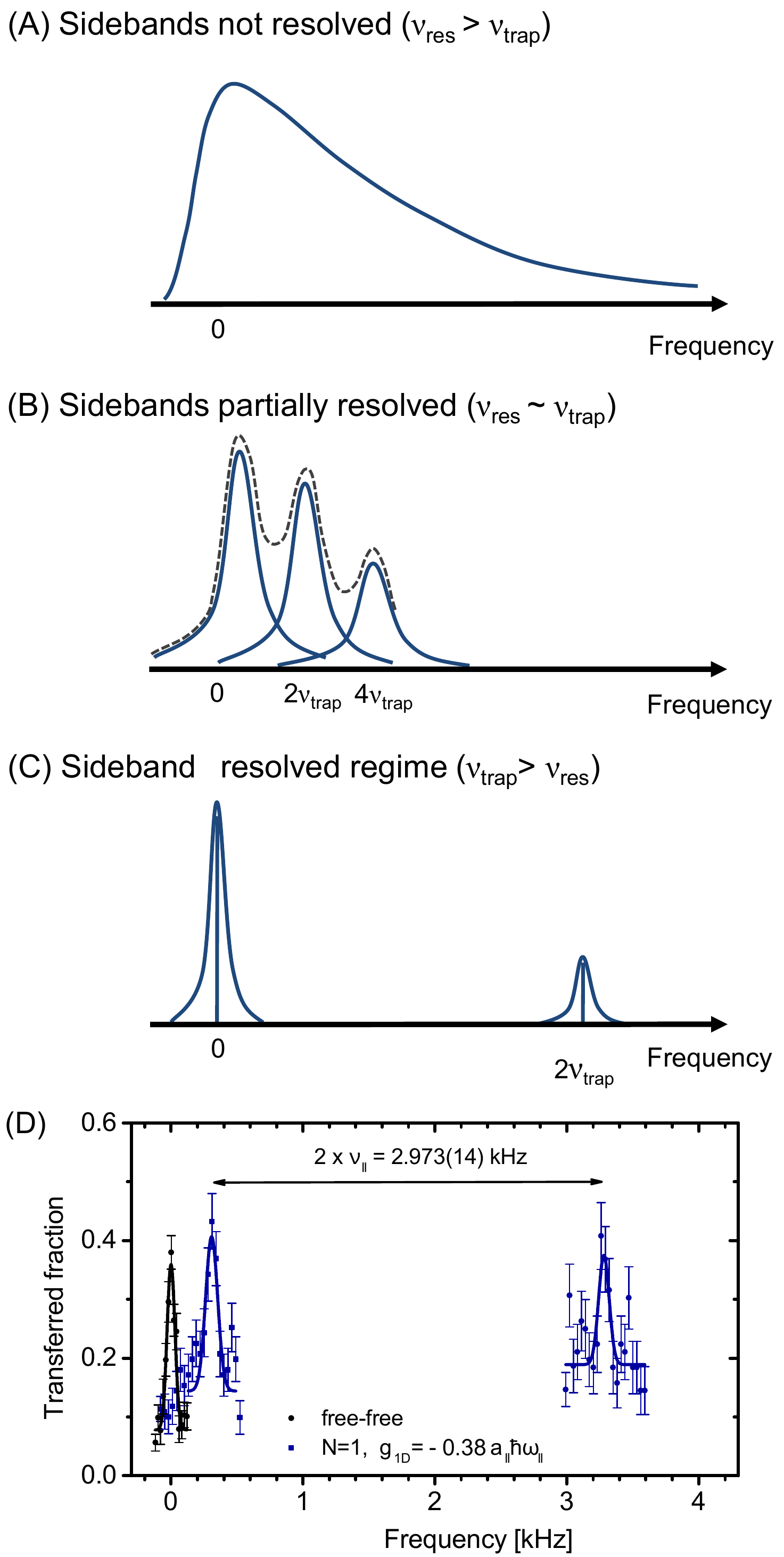}
	\label{fig:SOMfig3}
	\end{minipage}
	\begin{minipage}[b]{8.5cm} 
	\begin{itemize}
\item[]{
Figure S2: Different regimes of RF spectroscopy of a trapped sample. Panels (A-C) illustrate how the RF spectrum changes as a function of the RF resolution ($\nu_{\text{res}}$). In the resolved sideband regime (C) the peak position is insensitive to pulse duration and driving power. (D) RF spectrum in the resolved sideband regime for an attractively interacting two-particle system. In this measurements we drive from an interacting state to non-interacting states \cite{Zuern-thesis}. Depending on the applied RF frequency we either drive the transition to the non-interacting ground state (left blue peak) or to the next accessible excited state which is separated by $2\nu_\parallel$ (right blue peak). To resolve this sideband we increased the power of the RF pulse by 16dB and the duration of the pulse by a factor of $10$.}
\end{itemize}
	\hfill
	\end{minipage}
\end{figure}

To determine the free-free transition frequency $\nu_0$ we measure the fraction of transferred atoms as a function of the frequency of the applied RF pulse. To do this we prepare spin polarized samples of about $8$ atoms in state $\vert 2 \rangle$, apply the RF pulse, remove all atoms which are not in state $\vert 2 \rangle$ from the trap and count the number of remaining atoms \cite{Serwane2011}. The width of this transition -- and therefore the resolution of our energy measurement -- is determined by the stability of the magnetic offset field and the Fourier limit of the RF pulse.  As described in \cite{zuern2013} the magnetic field stability of our experimental setup is $\leq 5\,$mG, which corresponds to a frequency shift of about $5\,$Hz for the $\vert1\rangle$-$\vert2\rangle$ transition and about $50\,$Hz for the $\vert2\rangle$-$\vert3\rangle$ transition. To reduce decoherence and dephasing effects we choose a short RF pulse with a duration of $12.5\,$ms. This results in a total measured FWHM of $\lesssim 110\,$Hz. The power of the RF pulse is chosen such that we transfer about $50\%$ of the population to the final state. Achieving a transferred fraction of $50\%$ for the single impurity in an interacting system requires about $3\,$dB more RF-power.

To measure the frequency shift for the weakly interacting system we start from an interacting $\vert1\rangle$-$\vert3\rangle$ system at $B=589.69\,$G ($g_{\text{1D}} = 0.36$), drive the minority particle from state $\vert 1\rangle$ to state $\vert 2 \rangle$ and then detect the number of transferred minority atoms using the same method as described above. We take spectra for different numbers of majority atoms and extract the transition frequency from a Gaussian fit. To check for slow drifts of the magnetic field we measure the free-free transition before and after each measurement of the interaction energy and use the weighted mean of both free-free transition frequencies.  The error of the interaction shift is then determined by quadratic addition of the errors of the Gaussian fits to the spectra and the standard error of the mean of the two measurements of the free-free transition \cite{zuern2013}. The uncertainty of the interaction strength is dominated by the error of the transversal harmonic oscillator length $a_\perp$. Additionally, the interaction strength of the final state is not exactly zero ($g_{\text{1D}} = 0.006$) at the chosen magnetic field of $589.69\,$G and hence we add this difference to the error of $g_{\text{1D}}$.  Compared to these errors the influence of the uncertainty of the magnetic field on $g_{\text{1D}}$ is negligible

To obtain the interaction energy of a strongly interacting system we perform a measurement at a magnetic field of $B=634.51\,$G where a $\vert2\rangle$-$\vert3\rangle$ system has the same interaction strength as the 
$\vert 1\rangle$-$\vert3\rangle$ system at $B=589.69\,$G. This allows us to determine the interaction energy of the $\vert1\rangle$-$\vert3\rangle$ system at $B=634.51\,$G (corresponding to $g_{\text{1D}} = 2.80$) by measuring the energy difference of the $\vert1\rangle$-$\vert3\rangle$  and $\vert2\rangle$-$\vert3\rangle$  systems and adding the previously determined interaction energy of the weakly interacting system. An illustration of this two-step process is given in Fig. S3.  For this measurement the error of $g_{\text{1D}}$ is given by the sum of the errors of $g_{\text{1D}}$ resulting from the uncertainty of the transversal harmonic oscillator length and the small mismatch between the coupling strengths of the weakly interacting systems which we use as initial and final states.  

To determine the interaction energy of a system with intermediate interaction ($g_{\text{1D}}= 1.14$) we perform a similar two-step process indicated by the orange arrows in Fig. S3. An overview of all used RF-transitions is given in table S4.

\begin{figure}[htb!]
	\begin{minipage}[b]{8.5cm}
	\includegraphics[width=\textwidth]{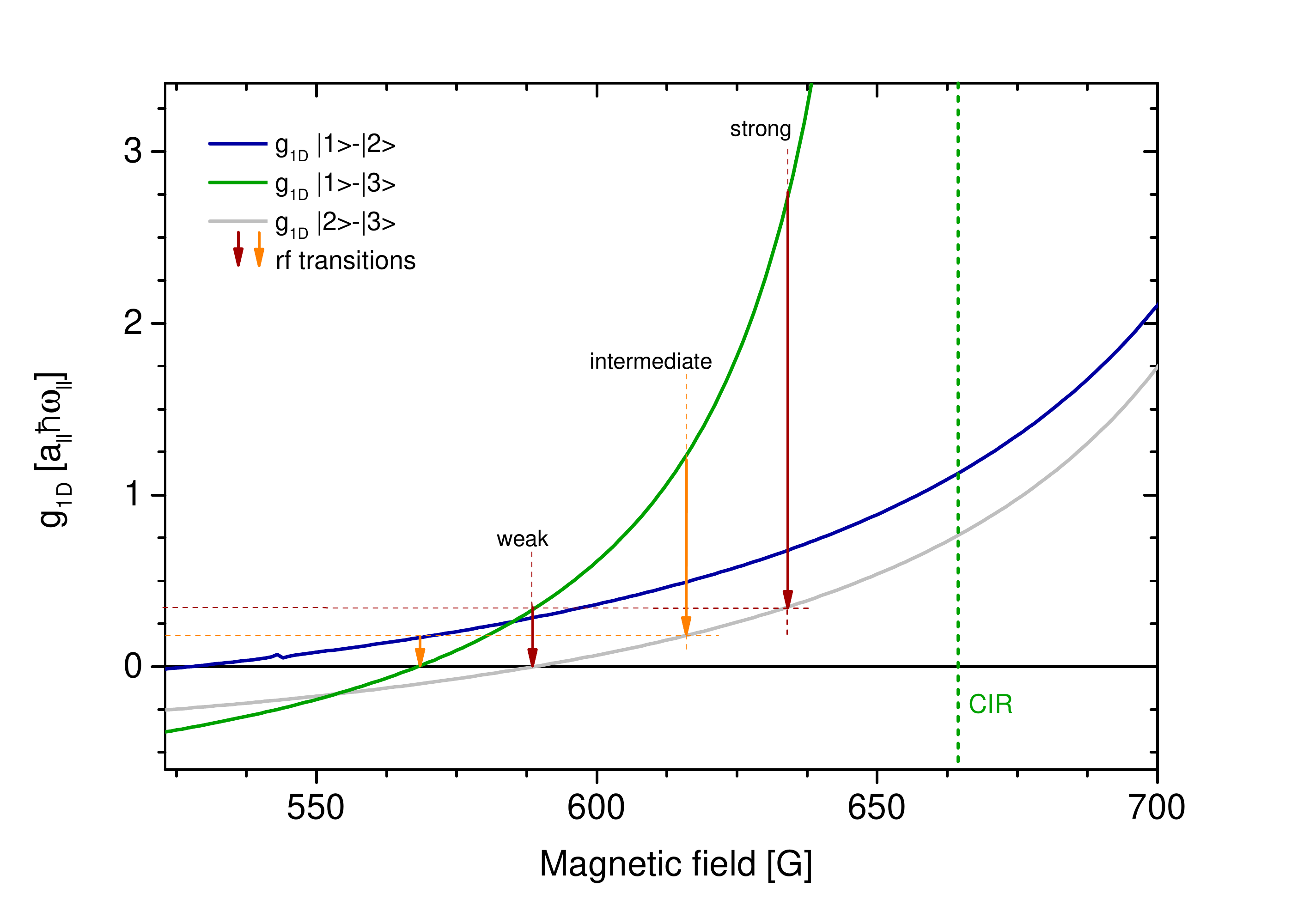}
	\label{fig:SOMfig2}
	\end{minipage}
	\begin{minipage}[b]{8.5cm} 
	\begin{itemize}
\item[]{Figure S3: One dimensional coupling constants $g_{\text{1D}}$ for all used hyperfine combinations as a function of  magnetic field. The green dashed line indicates the confinement-induced resonance (CIR) for the $\vert 1 \rangle$-$\vert 3 \rangle$ channel \cite{Bergeman2003, zuern2013}. The arrows indicate the RF transitions used for the determination of the interaction energies.
	 }
\end{itemize}
	\hfill
	\end{minipage}
\end{figure}

\begin{figure}[th]
	\begin{minipage}[b]{8.5cm}

  \begin{tabular}{ l | c |c |c| c|c|c}
                   & B [G]           & $g_{\text{1D},i}$  & $g_{\text{1D},f}$ & $\mid \uparrow \rangle$ & $\mid \downarrow_i \rangle$& $\mid \downarrow_f \rangle$\\ 
\hline \hline
	weak             	& 589.69 	  & 0.36  & 0.01 	& $\vert 3 \rangle$ 	& $\vert 1 \rangle$	&  $\vert 2 \rangle$ \\ \hline
	intermed. first  	& 568.30 		& 0.17 	& 0.00 	& $\vert 1 \rangle$ 	& $\vert 2 \rangle$	&  $\vert 3 \rangle$ \\ 
	intermed. second 	& 614.03 		& 1.14 	& 0.17 	& $\vert 3 \rangle$ 	& $\vert 1 \rangle$	&  $\vert 2 \rangle$ \\ \hline
	strong first     	& 589.69 		& 0.36 	& 0.01 	& $\vert 3 \rangle$ 	& $\vert 1 \rangle$	&  $\vert 2 \rangle$ \\ 
	strong second    	& 634.51 		& 2.80 	& 0.36 	& $\vert 3 \rangle$ 	& $\vert 1 \rangle$	&  $\vert 2 \rangle$ \\ \hline
	
\end {tabular}
	\label{somtable1}
	\end{minipage}
	
	\begin{minipage}[b]{8.5cm} 
	\begin{itemize}
	\item[]{}
\item[]{Table S4: List of all used rf-transitions. Given are the magnetic field values, coupling strengths in the initial and final system as well as the hyperfine states of the atoms in the initial and final system.}
\end{itemize}
	\hfill
	\end{minipage}
\end{figure}

\subsection*{Influence of the finite preparation fidelity on the measured spectra}
In addition to the system of interest also incorrectly prepared systems can lead to a spectroscopic signal which could cause systematic errors.
For strong interactions, these signals can be clearly distinguished from the actual peaks as their separation in energy is significantly higher than the width of these peaks (see Fig. 2 B). Therefore they have no influence on the determination of the interaction energies.\\
For weak interactions this condition is not fulfilled and one has to examine two relevant sources of imperfect preparations at this interaction strength. The first is the finite preparation fidelity of the initial sample, the second is the finite transfer efficiency of the RF Landau-Zener passage.\\
If the initial preparation has failed which occurs in $\lesssim12\%$ of the cases this will lead the appearance of additional peaks in the spectroscopic signal.
To estimate an upper bound for the influence of these peaks on the determination of the interaction energy we consider all imperfectly prepared systems to be $($$N$-$1$$)$-particle systems. Then we fit the measured spectra with a double-Gaussian lineshape. In this parametrization, the widths of both Gaussians are identical, the relative height of the Gaussians are fixed to the ratio between correctly and incorrectly prepared systems and the center of the second Gaussian is fixed by the energy determined for the $($$N$-$1$$)$-system. 
We find that the shift due to finite preparation fidelity is smaller than $5$\,Hz which is much smaller than the measured interaction shifts and can therefore be neglected.\\
The second process that lead to incorrectly prepared systems is the RF Landau-Zener passage which creates $\vert 1 \rangle$-$\vert 3 \rangle$ samples by converting the $N$ majority atoms from state $\vert 2 \rangle$ to state $\vert 3 \rangle$. The preparations that fail in this step due to the finite transfer efficiency ($t_z$=95\% per atom) can be divided into two categories: One where only a single state-$\vert 2 \rangle$ atom is not transferred and another one where more than one state-$\vert 2 \rangle$ atoms are not transferred. \\
In the first case which occurs with a probability of $N(1-t_z) \lesssim  25\%$ the state-$\vert 2 \rangle$ atom is assumed to undergo a 3-body loss involving the single impurity in state $\vert 1 \rangle$ and another state-$\vert 3 \rangle$ particle. The remaining systems then only consist of state-$\vert 3 \rangle$ particles and thus do not contribute to the measured spectrum as the RF addresses only the $\vert 1 \rangle$-$\vert 2 \rangle$ hyperfine transition.\\
The incorrectly prepared systems of the second category result from a second order process that involves at least two not transferred particles which has a probability on the order of $N(1-t_z)^2 \lesssim  12.5\%$. 
After a 3-body loss only atoms in state $\vert 2 \rangle$  and $\vert 3 \rangle$  remain. Since these atoms are randomly distributed over different trap states a nearly constant background is expected resulting from atoms in state $\vert 2 \rangle$. This offset does not considerably influence the determination of the interaction energy.

\subsection*{Corrections for anharmonicity and finite aspect ratio}
As described in the manuscript the trapping potential slightly deviates from a harmonic potential
and the Fermi energy of the majority component is determined by the energy of the last unoccupied trap level.
The energy levels of our trap have been determined by measuring the level spacing of the lowest trap levels and WKB calculations \cite{zuern2012}. 
The resulting $E_{F,\text{trap}}$ for different number of majority atoms is given in table S5.

	\begin{figure}[thb!]
	\begin{minipage}[b]{8.0cm}

\begin{tabular}{ l | c |c| c|c|c}
  N & 1 & 2 & 3 & 4& 5\\ \hline \hline
$E_{F,\text{harm}}[\hbar \omega_\parallel]$ & 1   &\, \;2\; \, &  3  &  4 & 5\\
$E_{F,\text{trap}}[\hbar \omega_\parallel]$ & 1.01   & \:2\:  &  2.97  &  3.93 & 4.88\\
		\end {tabular}
			\label{somtable2}
	\end{minipage}
	
	\begin{minipage}[b]{8.5cm} 
	\begin{itemize}
	\item[]{}
\item[]{Table S5: Comparison between the Fermi energies in a harmonic trap $E_{F,\text{harm}}$ and in our slightly anharmonic trap $E_{F,\text{trap}}$ for different number of majority atoms. We use the measured level spacing between the ground state and the second excited state of the trap as the reference for our harmonic approximation, since this is the most precisely determined level spacing of our trap \cite{zuern2012}.}
\end{itemize}
	\hfill
	\end{minipage}
\end{figure}
	
\begin{figure}[hbtp!]
\begin{minipage}[b]{8.5cm}

 \begin{tabular}{ l | c |c| c|c|c}
  N & 1 \cite{Busch1998, Idziaszek2004} & 2 \cite{Gharashi2012} & 3 \cite{Zuern-thesis} & 4& 5\\ \hline \hline
	$g_{\text{1D}}=0.36$ &  1 &  1 & 1 &1 & 1 \\
	$g_{\text{1D}}=1.14$ & 1.0051   & 1.0072  & 1.0072  & 1.0072  & 1.0072 \\
	$g_{\text{1D}}=2.80$ & 1.0106   & 1.0176  & 1.0206  & 1.0206& 1.0206\\
			\end {tabular}
			\label{somtable3}
	\end{minipage}
	
	\begin{minipage}[b]{8.5cm} 
	\begin{itemize}
	\item[]{}
\item[]{Table S6: Correction for the finite aspect ratio of our trap. Given is the correction factor by which we multiply our measured interaction energy to compare them to the true 1D theories \cite{Busch1998, Gharashi2012, Mcguire1965}.	For the weakly interacting system we neglect this correction because it is smaller than $2\times 10^{-3}$ which is below the resolution of our rf-spectroscopy. For intermediate (strong) interactions we have numerical results up to N=2 (N=3), which we use to calculate the ratio between the interaction energy of a true one-dimensional system and a system with a finite aspect ratio of $1:10$. For larger numbers of majority atoms we approximate the correction factor by the result of the $N=2$ ($N=3$) system.}
\end{itemize}
	\hfill
	\end{minipage}
\end{figure}	

Due to the finite aspect ratio of our quasi 1D trap our measured interaction energy deviates from the one in a true 1D system. We estimate this contribution by comparing the calculations for a quasi 1D system (aspect ratio $\omega_\parallel$:$\omega_\perp  \approx\,$1:10) and a true 1D system ($\omega_\parallel$:$\omega_\perp  \approx\,$1:$\infty$) which is shown in table S6.
For weak interactions the correction is significantly smaller than the error of the measurement of the interaction energy and hence is not considered. However, for our measurements at intermediate and strong interactions this contribution has been found to be as large as $2\%$ and can therefore not be neglected \cite{Busch1998, Idziaszek2004, Gharashi2012, Zuern-thesis}.
As we compare our measurement with theories for true 1D systems, we correct our data by the ratio between the quasi one-dimensional and the true one-dimensional system.
To our knowledge precise numerical calculations for both the true 1D and the quasi 1D system only exist for systems with a number of majority atoms up to $N=3$. From these calculations \cite{Gharashi2012,Zuern-thesis} we can determine a correction which is on the order of $2\%$ of the interaction energy for a strongly interacting system with $N=3$. 

For $N>3$ we expect the correction to be larger, however as can be seen from the corrections for strongly interacting systems for $N=1\dots 3$ in table S6 the increase of the correction as a function of particle number gets smaller for larger particle numbers.

Hence we approximate the relative correction for a strongly interacting system with $N>3$ by the relative correction of a system with $N=3$. As for lower interaction strength the corrections are smaller we can approximate the relative corrections for systems with intermediate interaction and $N>2$ by that of a system with $N=2$.

\footnotesize
	\begin{itemize}
		\item[]{
		\begin{center}
		\textbf{\_\_\_\_\_\_\_\_\_\_\_\_\_\_\_\_\_\_\_}
		\end{center}
		}
		\item[{[6]}]{\footnotesize F. Serwane, et al., Science \textbf{332}, 336 (2011).}
		\item[{[8]}]{\footnotesize T. Bergeman, M. G. Moore, M. Olshanii, Phys. Rev. Lett. \textbf{91}, 163201 (2003).}
		\item[{[9]}]{\footnotesize T. Busch, B.-G. Englert, K. Rzaewski, M. Wilkens, Foundations of Physics \textbf{28}, 549 (1998).}
		\item[{[10]}]{\footnotesize G. Z\"urn et al.,   Phys. Rev. Lett. \textbf{108}, 075303, (2012).}
		\item[{[11]}]{\footnotesize A. Schirotzek, C.-H. Wu, A. Sommer, M. W. Zwierlein, Phys. Rev. Lett. \textbf{102}, 230402 (2009).}
		\item[{[12]}]{\footnotesize Marco Koschorreck, et al., Nature \textbf{485}, 619622 (2012).}	
		\item[{[13]}]{\footnotesize C. Kohstall, et al., Nature \textbf{485}, 615618 (2012).}		
		\item[{[15]}]{\footnotesize J. B. McGuire, J. Math. Phys. \textbf{6}, 432 (1965).}
		\item[{[17]}]{\footnotesize G. Z\"urn, et al., Phys. Rev. Lett. \textbf{110}, 135301 (2013).}
		\item[{[19]}]{\footnotesize S. E. Gharashi, K. M. Daily, D. Blume,
		Phys. Rev. A \textbf{86}, 042702 (2012).}
		\item[{[27]}]{\footnotesize Z. Idziaszek, T. Calarco, Phys. Rev. A \textbf{71}, 050701 (2005).}
		\item[{[28]}]{\footnotesize G. Z\"urn, PhD thesis, Universit\"at Heidelberg (2012).}				
		\item[{[29]}]{\footnotesize P. O. Bugnion, G. J. Conduit, Phys. Rev. A \textbf{87}, 060502 (2013).}
		\item[{[37]}]{\footnotesize S. Sala, et al., Phys. Rev. Lett. \textbf{110}, 203202 (2013).}
		\item[{[38]}]{\footnotesize Note that in [10] the harmonic oscillator length $a_{\parallel}$ was defined differently and therefore the numerical value of $g_{\text{1D}}$ differs by a factor of $\sqrt{2}$.}					
  \end{itemize}
	\hfill


\begin{thebibliography}{38}
\bibitem[{\citenamefont{{The dimensionless interaction parameter $\gamma$ is
  also known as the Lieb Liniger
  parameter}}()}]{footnotedimensionlessparameter}
\bibinfo{author}{\bibnamefont{{The dimensionless interaction parameter $\gamma$
  is also known as the Lieb Liniger parameter}}}.
\bibitem[{\citenamefont{{Because the Fermi energy is also the only remaining
  energy scale in a 3D system we expect a similar convergence to a universal
  value determined by the Fermi energy times a numerical
  constant}}()}]{footnotenumericalconstant}
\bibinfo{author}{\bibnamefont{{Because the Fermi energy is also the only
  remaining energy scale in a 3D system we expect a similar convergence to a
  universal value determined by the Fermi energy times a numerical constant}}}.
\end{thebibliography}
\end{document}